\documentclass[a4paper]{article}

\usepackage{INTERSPEECH2021}
\usepackage{color}

\title{How do Voices from Past Speech Synthesis Challenges Compare Today?}
\name{Erica Cooper, Junichi Yamagishi}
\address{
  National Institute of Informatics, Japan}
\email{ecooper@nii.ac.jp, jyamagis@nii.ac.jp}

\begin{document}

\maketitle
\begin{abstract}
  Shared challenges provide a venue for comparing systems trained on common data using a standardized evaluation, and they also provide an invaluable resource for researchers when the data and evaluation results are publicly released.  The Blizzard Challenge and Voice Conversion Challenge are two such challenges for text-to-speech synthesis and for speaker conversion, respectively, and their publicly-available system samples and listening test results comprise a historical record of state-of-the-art synthesis methods over the years. In this paper, we revisit these past challenges and conduct a large-scale listening test with samples from many challenges combined.  Our aims are to analyze and compare opinions of a large number of systems together, to determine whether and how opinions change over time, and to collect a large-scale dataset of a diverse variety of synthetic samples and their ratings for further research.  We found strong correlations challenge by challenge at the system level between the original results and our new listening test.  We also observed the importance of the choice of speaker on synthesis quality.  
  % max 200 words.
\end{abstract}

\section{Introduction}

Since 2005, the annual Blizzard Challenge (BC) has provided researchers with a venue to compare their methods using common datasets and standardized evaluations.  Likewise, since 2016, the biennial Voice Conversion Challenge (VCC) has done the same for the task of speaker conversion.  Since the inception of the Blizzard Challenge, speech synthesis technology has transformed immensely, progressing through a diverse range of methods from unit selection synthesis, hidden Markov model based synthesis, and hybrid models to present-day state-of-the-art approaches such as end-to-end neural network based synthesis.  In recent years, speech synthesis technology has also reached an overall level of acceptability to the general public where it is now very commonly used in various everyday consumer technologies.

In addition to providing shared data and evaluations for researchers to compare their approaches, the Blizzard and Voice Conversion Challenges also make the synthesized samples and raw listening test results publicly available, which is an invaluable resource for studying different models and approaches over time.  Nevertheless, it is well-known that results from different listening tests cannot be meaningfully compared to each other \cite{hossfeld2016qoe} because the setting and conditions of the tests are not identical -- the set of systems is different, and in particular the differing best and worst systems each year provide listeners with a completely different context for their evaluations.  For this reason, we have gathered samples from past Blizzard and Voice Conversion Challenges into one new large-scale listening test which enables us to compare many past text-to-speech and voice conversion systems together.  This allows us to answer the following research questions:

\begin{itemize}
    \item How reproducible are MOS test results? 
    \item How do past listening test results compare to ratings gathered in the present day?
    \item Will results still correlate even though the listening test context has changed?
    \item Can we observe the improvement of speech synthesis technology over the years in this data?
    \item How does quality of text-to-speech synthesis and voice conversion compare?
    \item What is the effect of the target speaker data on perceived synthesis quality?
\end{itemize}

Furthermore, a dataset of many years of synthesized samples along with their ratings from a single listening test will be a useful resource for training automatic evaluation metrics such as MOSNet \cite{Lo2019}.  In this paper, we will describe the design of a large-scale listening test that aims to compare quality of a diverse range of synthesis methods from past years' Blizzard and Voice Conversion Challenges.  We will then present what we learned from this test in terms of synthesized speech and listener preferences of natural speech.  To the best of our knowledge, this is the first time that samples from different years' challenges, as well as a combination of both text-to-speech synthesis and voice conversion samples, have been compared in one listening test together.  

\section{Related Work}

In a 2014 overview of a decade of past Blizzard challenges \cite{king2014measuring}, it was observed that unit selection based systems consistently had the best naturalness ratings over the years, whereas statistical parametric methods such as hidden Markov model based synthesis produced the most intelligible speech.  Hybrid systems were beginning to show signs of incorporating the best of both worlds.  While it was noted that ``naturalness'' as a basis for rating speech audio is inherently poorly-defined, the consistency of listener judgments shows that listeners are nevertheless able to understand and complete the task.  In another meta-study \cite{hinterleitner2013intelligibility}, nine different past studies of human ratings of synthetic speech revealed five common aspects (naturalness, prosodic quality, intelligibility, disturbances, and calmness) that were consistently salient to listeners' judgements.  There have also been a number of studies that re-visit or reproduce listening tests in order to study the reliability of MOS tests.  For instance, \cite{naderi2020open} ran the same listening test both in lab and as an online crowdsourced task and found strong correlations between ratings in both settings, and furthermore ran the crowdsourced test five times on five different days with five different sets of listeners and also found good reliability between the sets of results.  \cite{zequeira2019intra} also found good agreement and strong correlations between an in-lab listening test and a crowdsourced one.  In a 2015 re-visitation of the 2013 Blizzard Challenge results, \cite{wester2015we} studied the stability of the significant differences between systems, finding that the results stabilize and have good reliability and discriminative power when at least 30 different listeners are included in the test.

Due to the expense and time-consuming nature of conducting subjective listening tests, there has long been interest in the development and use of objective measures for evaluating synthesis quality, and in particular, with the recent advances in neural network based modeling approaches, past listening test results can be used to train models for this purpose.  For example, MOSNet \cite{Lo2019} trained an end-to-end model for naturalness assessment on the VCC 2018 listening test results to predict human ratings of voice-converted speech.  They further extended their model to predict speaker similarity in addition to MOS.  While they found high correlations at the system level but only fair correlations at the utterance level due to large variances between listeners, \cite{leng2021mbnet} extended MOSNet to learn from this listener variation by incorporating a listener bias network that takes the listener label into consideration.  In addition to improving utterance-level correlations when the appropriate listener label is given, overall system-level correlations were also improved.  Another extension of MOSNet was conducted in \cite{williamscomparison}, in which the models trained for VC were found not to generalize well to TTS, so MOSNet models were trained on the ASVspoof 2019 Logical Access dataset \cite{Todisco2019}, which contains synthesized speech from 13 different speech synthesizers and voice conversion systems trained on the same set of speakers.  Eight different feature representations were studied to determine which one is best for this type of evaluation task.  While \cite{ITUMOS} cautions that even an objective measure depends on its context (i.e.,~its training data) much in the same way that human listening tests do, it is our hope that very large-scale listening test data such as that collected in our study will provide sufficient context to train objective measures that have good generalization capability in the future.

\section{Listening Test Design}

We gathered samples and ratings from past Blizzard and Voice Conversion Challenges\footnote{https://www.cstr.ed.ac.uk/projects/blizzard/data.html}.  We focused on English-language synthesis and the main Hub tasks for each year.  The Blizzard Challenge years that we included were 2008 \cite{blizzard2008}, 2009 \cite{blizzard2009}, 2010 \cite{blizzard2010}, 2011 \cite{blizzard2011}, 2013 \cite{blizzard2013}, and 2016 \cite{blizzard2016}, as well as all Voice Conversion Challenge years (2016 \cite{vcc2016description,vcc2016analysis}, 2018 \cite{vcc2018}, and 2020 \cite{vcc2020,Rohan2020}).  We also included samples from a number of systems from ESPnet \cite{watanabe2018espnet}, which is a popular open-source toolkit for end-to-end speech and language technologies, since samples for a number of implemented text-to-speech architectures have been released along with their listening test results \cite{hayashi2019espnettts}.  Our total number of systems, including natural speech, was 187.

We chose 38 samples for each of the 187 systems, balancing where relevant over genre (e.g. news, audiobook, conversational).  We excluded semantically-unpredictable sentences, which were used in past Blizzard challenges mainly for intelligibility evaluation, as well as any other genres which were not included in the original naturalness evaluations, and genres for which there were no corresponding natural speech samples.  For voice conversion systems, we balanced over all combinations of source and target speakers.  Even though VCC 2020 had both intra-lingual and cross-lingual tasks, we only included samples from the intra-lingual task.  Some challenges did not have 38 unique test utterances, so in those cases we included repeat samples.  To avoid differing sampling rates as a confounding factor, we downsampled all audio to 16kHz, and conducted amplitude normalization using sv56 \cite{sv56}.

Each listening test set consisted of one sample from each of the 187 different systems.  Listeners could listen to each sample as many times as they liked, but were required to play the entire sample at least once and choose a rating for it before proceeding to the next one.  Listeners were asked to rate each sample on a 5-grade Mean Opinion Score (MOS) scale from 1 (very bad) to 5 (very good).  In order to get ratings from as many different listeners as possible, each listener was only permitted to evaluate one set.  Each set was rated by eight different listeners, and overall, 304 different listeners participated in our test.  Due to the constraints of our location, we recruited Japanese native listeners to participate in our test, but we also note the very strong correlations with native English listeners reported in \cite{vcc2020}.  Listener gender demographics were 141 male, 159 female, and 4 other.  Listener age demographics were 48 listeners between 18 and 29 years old, 118 listeners in their 30s, 90 listeners in their 40s, 35 listeners in their 50s, 12 listeners in their 60s, and one listener age 70 or older.  We measured significant differences between systems using the Mann-Whitney U test, following \cite{Rosenberg2017}, at a level of p$<$0.05, with Bonferroni correction for multiple comparisons.

\section{Results and Analysis}

\begin{figure*}[!t]
\centering
\includegraphics[width=1.0\textwidth]{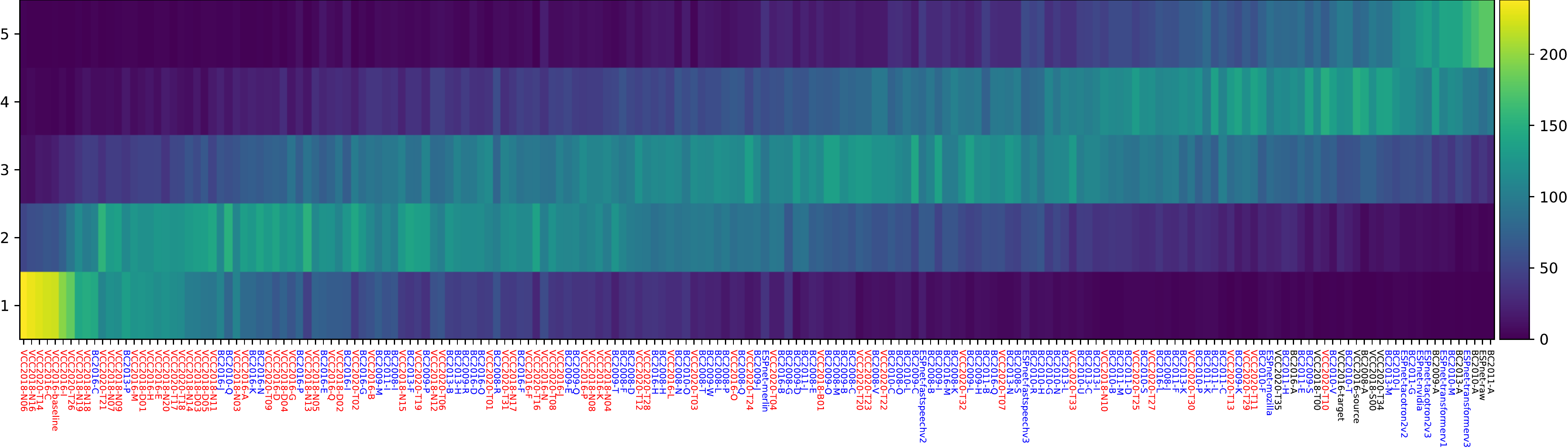}
\caption{Histogram of MOS ratings for 187 systems. Natural speech system names are indicated in black text, TTS systems are \textcolor{blue}{blue}, and voice conversion systems are \textcolor{red}{red}.} 
\label{fig:hist}
\end{figure*}

A histogram of the ratings for all 187 systems, arranged from lowest to highest MOS, can be seen in Figure \ref{fig:hist}.  We found moderate listener agreement, with both Krippendorff's alpha and intra-class correlation equal to 0.50.

\begin{figure*}[!t]
\centering
\includegraphics[width=1.0\textwidth]{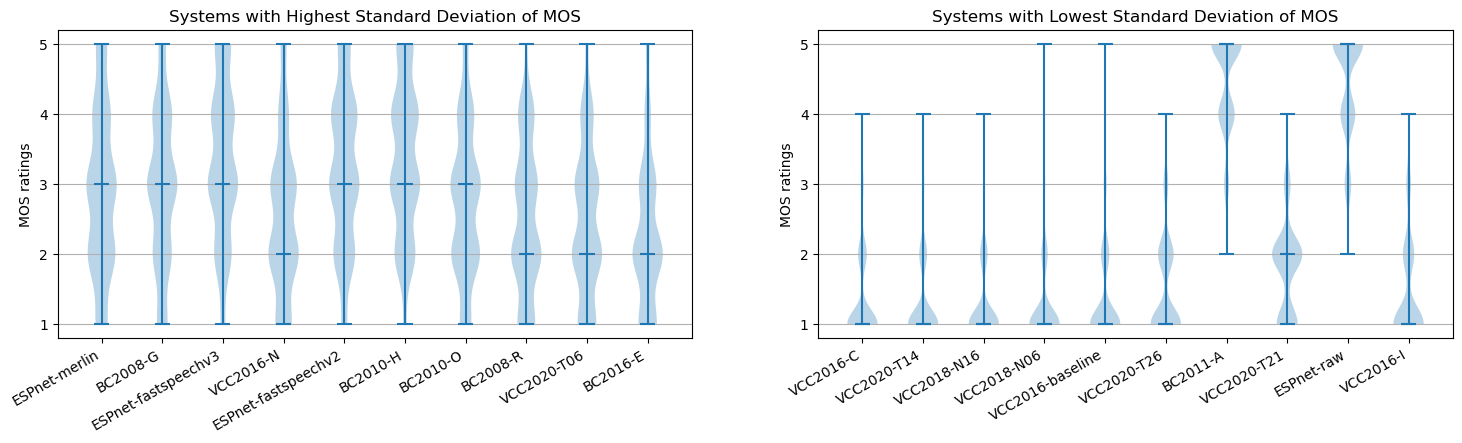}
\caption{Violin plots of the systems with the highest and lowest standard deviations.} 
\label{fig:violin}
\end{figure*}

Looking at the standard deviations of each system, we noticed that some systems were less agreed-upon than others.  ESPnet-Merlin, a DNN-based parametric model trained using the Merlin toolkit \cite{merlin}, had the highest standard deviation, with an almost equal number of 5 and 1 ratings.  The systems with the lowest standard deviations tended to be natural speech (very highly rated) or the lowest-rated systems.  Violin plots of the rating distributions of the most- and least-agreed-upon systems are in Figure \ref{fig:violin}.

\subsection{Best and Worst Systems}

Systems are named according to the challenge that they came from, followed by the team letter name or other system identifier.  The five best synthesized systems, which were not significantly different from one another, are the following:
\begin{itemize}
    \item ESPnet-transformerv3
    \item BC2010-M
    \item ESPnet-transformerv1
    \item ESPnet-tacotron2v3
    \item ESPnet-nvidia
\end{itemize}

It is notable that four out of the five best-rated systems are from ESPnet.  One drawback of our listening test as compared to the standard Blizzard evaluations is that we are mixing systems that were trained on a variety of different databases, so it becomes more difficult to determine whether a model is inherently better or if listeners simply prefer the sound of the voice data on which it was trained. We will discuss this more in Section \ref{sec:naturalspeech}.

%It is somewhat surprising that LJSpeech was in fact so strongly preferred -- these are relatively high-quality recordings of read audiobooks, but it is nevertheless a kind of ``found'' data from the LibriVox project\footnote{https://librivox.org} and the audio contains audible reverberation and compression artifacts.  Despite this, listeners still significantly preferred this speech over e.g. the Blizzard 2013 audiobook data, which was professionally-recorded.

The group of worst systems which are not significantly different from one another are as follows:
\begin{itemize}
    \item VCC2018-N06 
    \item VCC2018-N16
    \item VCC2020-T14
    \item VCC2016-C
    \item VCC2016-baseline
\end{itemize}

Text-to-speech and voice conversion systems are rarely compared together in the same listening test, but this large-scale test gave us the opportunity to do so.  It is notable that the worst-rated systems are all voice conversion ones.  Is the state of the art of text-to-speech synthesis better overall (in terms of naturalness) than that of voice conversion?  One consideration is that voice conversion from a source speaker to a target speaker of a different gender may produce worse speech signal quality than the same-gender condition, since the distance between source and target speaker is farther.  So, we tried excluding samples where the source and target speakers were different genders and re-computed MOS.  Although the ordering changes slightly, we find that the worst systems are still voice conversion ones.  Furthermore, although the MOS values tend to improve slightly by only considering same-gender conversion, we find that it is generally not statistically significant -- only four out of the 73 voice conversion systems show any significant improvement.  Since we have both TTS and VC systems from 2016, we can also compare the best systems from both challenges in the same year:  BC2016-L was rated as significantly better than VCC2016-O.

%%%%%%%%%%  EC: taking this out to reclaim some space
%%%%% and also a reviewer said this VC part was too long.
%To further compare TTS and VC, we can make the following observations:

%\begin{itemize}
%    \item The overall best-rated TTS system (ESPnet-transformerv3) is significantly better than the best VC system (VCC2020-T10).
%    \item The best-rated TTS system from a Blizzard challenge (BC2010-M) is also significantly better than VCC2020-T10.
%    \item The worst-rated VC system (VCC2018-N06) is significantly worse than the worst TTS system (BC2016-C).
%    \item We have both TTS and VC systems from 2016, so we can compare the best systems from both challenges in the same year:  BC2016-L is significantly better than VCC2016-O.
%\end{itemize}

Another consideration is that the Voice Conversion Challenges provide teams with much less data per speaker, often as few as around 80 utterances, whereas Blizzard Challenge data is typically on the order of a few hours or thousands of utterances.  These kinds of low-resource data conditions make it more challenging to achieve a high level of naturalness, which is apparent from the listener ratings.

\subsection{Correlations with past challenge results}

At the system level, challenge by challenge, we found very strong correlations, using both the Pearson correlation coefficient (PCC) and the Spearman rank correlation coefficient (SRCC), between the original listening test results and the new ones.  We report these values and also root mean squared error (RMSE) in Table \ref{tab:correl}.  At the utterance level, we find lower but still moderately positive correlations.  Individual utterance-level scores were not available for BC2013 and BC2016.
%% these high correlations were surprising in a good way.

\begin{table}[th]
\scriptsize
  \caption{System-level and utterance-level PCC, SRCC, and RMSE between original and new listening test results by challenge or set of systems}
  \label{tab:correl}
  \centering
  \begin{tabular}{ l  l  l  l  l l l }
    \toprule
     & \multicolumn{3}{c}{\textbf{System-level}} & \multicolumn{3}{c}{\textbf{Utterance-level}} \\
    \textbf{Challenge} & \textbf{PCC} & \textbf{SRCC} & \textbf{RMSE} &  \textbf{PCC} & \textbf{SRCC} & \textbf{RMSE} \\
    \midrule
    BC2008   & 0.93 & 0.89 & 0.33  & 0.70 & 0.67 & 0.62  \\
    BC2009   & 0.97 & 0.95 & 0.48  & 0.76 & 0.72 & 0.64 \\
    BC2010   & 0.93 & 0.98 & 0.66  & 0.74 & 0.73 & 0.85 \\
    BC2011   & 0.91 & 0.90 & 0.76  & 0.76 & 0.67 & 0.87 \\
    BC2013   & 0.97 & 0.98 & 0.49  & - & - & - \\
    BC2016   & 0.97 & 0.93 & 0.40  & - & - & - \\
    \midrule
    VCC2016  & 0.97 & 0.92 & 0.42  &  0.56 & 0.53 & 1.12 \\
    VCC2018  & 0.96 & 0.91 & 0.77  & 0.55 & 0.53 & 1.10 \\
    VCC2020  & 0.98 & 0.96 & 0.23  & 0.87 & 0.87 & 0.48 \\
    \midrule
    ESPnet   & 0.99 & 0.98 & 0.09  & 0.73 & 0.61 & 0.59    \\
    \bottomrule
  \end{tabular}
  
\end{table}

The large RMSE values show the effects of context -- even though year-by-year correlations are strong, the overall values of the ratings themselves compared to the original ones do vary.

\subsection{Improvements of speech synthesis over time}
Year by year, is the best system in each challenge better than the previous year's best system?  At what point in time did synthesized speech reach the quality where it was not rated as significantly different from natural speech?  Table \ref{tab:bestbyyear}  shows the MOS of the best system for each challenge, whether its MOS has improved over the previous challenge's best system, whether this difference is significant, and whether this challenge's best system is significantly different from that same year's natural speech.  We can observe that while VCC best systems do improve challenge by challenge, the best Blizzard Challenge system from 2016 was rated as significantly worse than the best system from BC2013.  This is likely due to the effects of the different training corpora.  We can also observe that in 2010 and onwards (excepting 2011) for TTS, and from 2018 onwards for voice conversion, the best systems' MOS ratings were not significantly different from natural speech.

\begin{figure}[!htbp] %[!t]
\centering
\includegraphics[width=1.0\columnwidth]{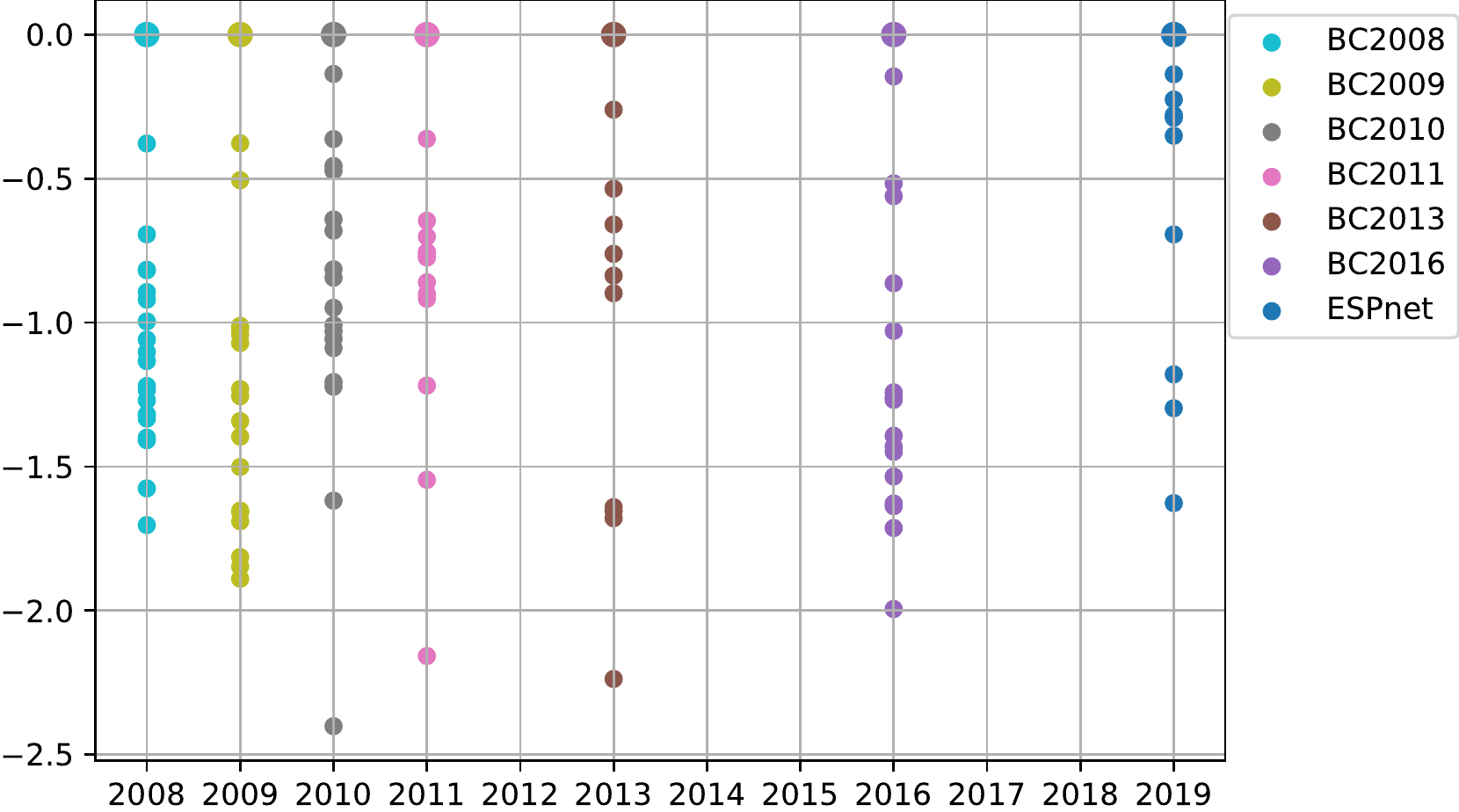}
\caption{Difference of each system from natural speech, computed from averaged z-score-normalized ratings by listener for each challenge.  The top row of larger dots are natural speech.} 
\label{fig:scatter}
\end{figure}

Since some listeners may have strong preferences about the speaker voice chosen for a given year's challenge, and since these preferences will therefore skew that listener's ratings for all systems trained on that dataset, adjusting for these preferences may allow us to see more clearly an overall trend of how TTS systems perform relative to the quality of natural speech over time.  Z-score normalization was conducted based on statistics of all of a listener's ratings for systems in a single challenge, normalized average scores were computed for each system from the normalized individual ratings, and differences were computed between the normalized score of a given year's natural speech and of each system from that year.  Results are plotted in Figure \ref{fig:scatter}.  We can see that the gap between natural speech and the best system becomes smaller year by year (with the exception of a very good best system in 2010), and also that a larger number of systems tend to approach natural speech over time.

\begin{table}[t]
\footnotesize
  \caption{Best system in each challenge compared to the previous challenge's best system and to natural speech -- whether MOS has improved since the last challenge (Impr.?), whether the difference is significant (Sig.?), and whether the difference in MOS to that year's natural speech is significant (Sig. (Nat)).}
  \label{tab:bestbyyear}
  \centering
  \begin{tabular}{lllll}
    \toprule
    \textbf{Year : Best system} & \textbf{MOS} & \textbf{Impr.?} & \textbf{Sig.?}  & \textbf{Sig. (Nat)} \\
    \midrule
    BC2008 : J & 3.63 & & & \checkmark \\
    BC2009 : S & 3.87 & \checkmark & x & \checkmark \\
    BC2010 : M & 4.27 & \checkmark & \checkmark & x \\ 
    BC2011 : G & 4.12 & x & x & \checkmark \\
    BC2013 : M & 4.01 & x & x & x \\
    BC2016 : L & 3.63 & x & \checkmark & x \\
    \midrule
    VCC2016 : O & 2.86 & & & \checkmark \\
    VCC2018 : N10 & 3.55 & \checkmark & \checkmark & x \\
    VCC2020 : T10 & 3.88 & \checkmark & x & x \\
    \midrule
    ESPnet : transformerv3 & 4.33 & & & x \\
    \bottomrule
  \end{tabular}
\end{table}

\subsection{Objective measures}

We objectively measured all of the samples included in our listening test using a number of metrics: word error rate (WER) using the IBM Watson speech-to-text API\footnote{https://www.ibm.com/cloud/watson-speech-to-text}, signal-to-noise ratio (SNR) using the WADA SNR algorithm \cite{kim2008robust}, the ITU-T P.563 method for objective speech quality assessment \cite{p563}, and a pretrained MOSNet model from \cite{williamscomparison}.  Although there are many  pretrained MOSNet models to choose from, we chose this one because the fact that it was trained on a variety of TTS and VC systems from ASVspoof makes it a good match for our domain, and furthermore, other models which were trained on VCC data would not be valid to use since we would be testing on those models' training data.  Surprisingly we found that WER had the strongest (negative, as expected) Pearson's correlation with MOS at r=-0.52.  SNR had a weak negative corrlation of r=-0.17.  The p563 measure had a very weak correlation of r=0.05, and surprisingly, MOSNet had the weakest correlation of all at r=0.03.  There is clearly room for improvement in terms of generalizable objective measures for synthesized speech.

\section{Natural speech preferences and effects of the corpus on TTS}
\label{sec:naturalspeech}

In \cite{hinterleitner2014influence}, the voice of the speech corpus was found to have a significant effect on the ratings of the synthesized speech.  They caution that the selection of the speaker for the training corpus is crucial due to the large effects that the speaker can have on the perceived quality of the synthesized speech.  \cite{williamscomparison} similarly found that the speaker has a large effect on synthesis quality, and that systems trained on data from certain speakers reached a consistent quality, regardless of the type of synthesis model used.  We observe this in our listening test data as well.  This is a confounding factor that prohibits meaningful direct comparisons between systems from different challenges; however, for training a system such as MOSNet, it is important to be able to replicate these human preferences even if they are simply based on characteristics of the speaker data.

\subsection{Metadata}

%% TODO : I am not actually sure that some speakers were NOT professional speakers.....
We have useful metadata about various speaker characteristics, such as gender, dialect (American vs.~British), and whether or not the speaker is a professional voice talent (speakers who were not specifically stated to be professional speakers in the data descriptions were assumed not to be).  We found a significant preference for professional speech over non-professional speakers, a marginally significant preference for female speakers over male speakers at p=0.05, and no significant preference between British and American speakers.  The preference for professional speech may account for some of the difference between voice conversion and text-to-speech systems, since the voice conversion challenges rely on non-professional speech.
%% DAPS dosn't say whether they are prrofessional speakers or not,
%% EMIME:  "The bilingual talkers were recruited via the Edinburgh University Careers Services, the native English talkers were selected locally from CSTR and Informatics. Bilingual in this context means a person who has the ability to speak and read two languages."

According to \cite{dall2014rating}, listeners tend to rate spontaneous speech as more natural, even if not explicitly instructed to pay attention to style.  So, we consider whether the genre or style of the natural speech has an effect on perceived naturalness.  The three main genres that we included from the Blizzard Challenges are news, book sentences, and a ``conversational'' genre which is not spontaneous conversational speech, but rather meant to be speech from a virtual conversational agent whose purpose is to help the user search for restaurants and navigate the results.  We find that news sentences are overall rated the most natural with a MOS of 4.36 and the conversational genre had a MOS of 4.14.  The book sentences were rated as significantly less natural than the news sentences, with a MOS of 4.09.  It is surprising that the book speech was rated as less natural, but the highly expressive style of many of the book sentences may come across as unnatural out of context. 

One interesting observation we made during these analyses is that although the speech data came from the same speaker in both Blizzards 2008 and 2009, there was a significant preference for the Blizzard 2009 natural speech.  In fact, even controlling for genre by considering only news utterances, we still found a significant difference.  Listening to samples from these sets, we observed that the audio quality was much better for the 2009 samples.  From this we can conclude that listeners are able to consistently pick up on such differences in recording quality.
%% TODO: it may not have been a different recording condition but rather a different day when the speaker's voice quality was worse.

\subsection{Speaker characteristics}

We next consider whether there are certain speaker characteristics that listeners tend to favor when rating naturalness.  For each speaker, using Praat \cite{praat}, we measure the minimum, maximum, mean, and standard deviation of f0 and energy, as well as noise-to-harmonic ratio (NHR), jitter, and shimmer.  We found moderate negative correlations with MOS for shimmer (r=-0.46), NHR (r=-0.41), and mean energy (r=-0.37), and a moderate positive correlation for standard deviation of energy (r=0.41).  A study of vocal attractiveness \cite{bruckert2010vocal} also found that harmonic-to-noise ratio (the inverse of NHR) was significantly correlated to ratings of vocal attractiveness, suggesting that perceptions of naturalness and vocal attractiveness may be related.  Furthermore, \cite{lee2018comparison} observed that selecting speakers with low mean energy for training statistical parametric speech synthesis models resulted in more intelligible synthetic speech, which we have also observed correlates with better naturalness ratings.

\subsection{Effect of corpus on benchmark systems}

\begin{figure}[!t]
\centering
\includegraphics[width=1.0\columnwidth]{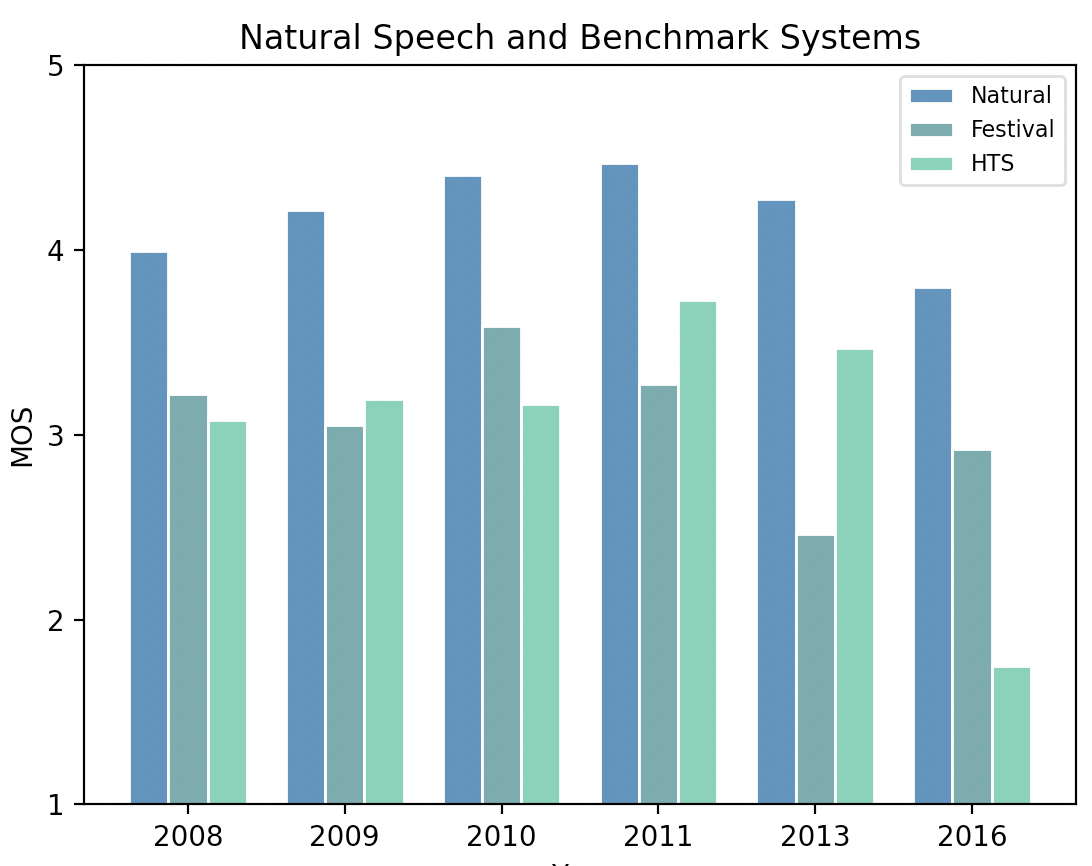}
\caption{MOS for natural speech and benchmarks for each Blizzard year} 
\label{fig:benchmarks}
\end{figure}

Every Blizzard evaluation contains samples from two benchmark systems: Festival \cite{festival} and HTS \cite{hts}.  The Festival benchmark system is the same every year, and the HTS benchmark is that year's HTS version.  Festival and HTS are denoted as systems B and C respectively in each Blizzard challenge.  The inclusion of these benchmarks allows us to study the effect of the speaker data on TTS systems that are mostly consistent.  Comparing Pearson and Spearman corrrelations between the scores for these benchmark systems each year and the corresponding natural audio, we find moderate correlations for Festival (Pearson r=0.33, Spearman r=0.54) and strong correlations for HTS (Pearson r=0.87, Spearman r=0.90).  This indicates that while both systems' synthesis output reflects preferences about the chosen speaker used for training, HTS is more sensitive to the choice of data. Groupings of the MOS of benchmark systems with their respective natural  speech can be seen in Figure \ref{fig:benchmarks}.

\section{Discussion and Future Work}

In a large-scale listening test combining samples from various Blizzard Challenges, Voice Conversion Challenges, and ESPnet models, we showed the reliability of MOS tests through their strong correlations with MOS results from past tests.  In doing so, we also produced a very large dataset of synthesized samples from 187 different systems, each with eight human ratings for naturalness, and all in the same listening test context, with both text-to-speech and voice conversion systems rated together, which can be used for further analysis and for training MOSNet-type systems for automatic objective evaluation.  We also observed the importance of the choice of speaker for the training data on synthesis quality, and identified some speaker characteristics for which listeners had preferences.  By adjusting for individual listener preferences and measuring distance to natural speech, we can observe the trend of improvement in TTS over time as more systems approach the quality of natural speech.

We have observed that some systems have clear agreement, whereas others, such as ESPnet-Merlin, have a wider distribution of scores.  For these such less-agreed-upon systems, it would be interesting to know the source of these disagreements and what makes them so ``controversial'' -- e.g., if certain types of artifacts or unnaturalness are very salient to some listeners but not others, or if the variation comes from large differences in quality of synthesis by utterance.

In future work, we will conduct a similar listening test with native English listeners, and also collect ratings for speaker similarity.  These large datasets will allow us to to train or fine-tune MOSNet models for this test context.  Having similar listening test data for both English and Japanese listeners will also enable us to study cross-cultural aspects of preferences for speaker characteristics and speaking styles.

\section{Acknowledgements}

\footnotesize{We would like to thank the organizers of the Blizzard Challenges and Voice Conversion Challenges for making the samples and listening test data freely available.  We would also like to thank Simon King and Alan Black for answering our questions about the data, Tomoki Hayashi for kindly providing utterance-level MOS ratings for the ESPnet systems, and S\'ebastien Le Maguer for helpful discussions. This study is supported by JST CREST Grants (JPMJCR18A6 and JPMJCR20D3) and MEXT KAKENHI Grants (21K11951), Japan.}
%This study is supported by JST CREST Grants (JPMJCR18A6 and JPMJCR20D3), MEXT KAKENHI Grants (18H04112, 21H04906, 21K11951, 21K17775), Japan, and Google AI for Japan program. Some of the experiments were conducted on TSUBAME 3.0 of Tokyo Institute of Technology.

\bibliographystyle{IEEEtran}

\bibliography{template}

\end{document}